\documentclass[twocolumn,preprintnumbers,amsmath,amssymb]{revtex4}
\usepackage{graphicx}% Include figure files
\usepackage{dcolumn}% Align table columns on decimal point
\usepackage{bm}% bold math

\newcommand{\beq}{\begin{equation}}
\newcommand{\eeq}{\end{equation}}
\newcommand{\bea}{\begin{eqnarray}}
\newcommand{\eea}{\end{eqnarray}}
\newcommand{\bdm}{\begin{displaymath}}
\newcommand{\edm}{\end{displaymath}}

\newcommand{\kt}{\textrm{\textbf{k}}}

\def\as{\alpha_s}

\def\ord{{\cal O}}

\def \msb{\overline{\textrm{MS}}}

\begin{document}
\preprint{MAN/HEP/2009/22} \preprint{Edinburgh  2009-07}
\title{Drell-Yan processes in the high-energy limit \footnote{Talk given at  ``DIS 2009'', Madrid, Spain, 26$^{\rm th}$-30$^{\rm th}$ Apr 2009 }}% Force line breaks with \\

\author{Simone Marzani}
 \email{simone.marzani@manchester.ac.uk}
 
 \affiliation{School of Physics \& Astronomy, University of Manchester,\\Oxford Road, Manchester, M13 9PL, England, U.K.}%Lines break automatically or can be forced with \\

 \author{Richard D.~Ball}
 \email{rdb@ph.ed.ac.uk}
 \affiliation{ School of Physics and Astronomy, University of Edinburgh,\\
Mayfield Road, Edinburgh EH9 3JZ, Scotland, U.K.}

%\author{Second Author}%

%\affiliation[Also at ]{%
%Authors' institution and/or address\\
%This line break forced with \textbackslash\textbackslash
%}%

%\author{Charlie Author}
 %\homepage{http://www.Second.institution.edu/~Charlie.Author}
%\affiliation{
%Second institution and/or address\\
%This line break forced% with \\
%}%

%\date{\today}% It is always \today, today,
             %  but any date may be explicitly specified

\begin{abstract}
We present the analytic computation of leading high-energy logarithms of the inclusive Drell-Yan and vector boson production cross-section.
 We also study the phenomenological relevance of the high-energy corrections for Drell-Yan processes at the LHC. We find that the resummation corrects the NNLO result  by no more than a few percent, for values of the invariant mass of the lepton pair below $100$~GeV.
\end{abstract}

\maketitle

\section{Drell-Yan and vector boson production cross-section at the LHC}

Accurate perturbative calculations of benchmark processes such as Drell-Yan production of a lepton pair and the closely related  vector boson production cross-section
are an essential component of the LHC discovery programme. Outstanding precision has been reached in the calculation of radiative corrections in perturbative QCD.

The inclusive cross-sections have been known at NNLO for a long time~\cite{Altarelli:1978id,Matsuura:1990ba,
Hamberg:1990np,Blumlein:2005im}. The calculation of the rapidity distribution at the same accuracy has been performed in~\cite{NNLOrap}.
More recently the fully differential cross section has also been computed.
The resummation of threshold logarithms
is known up to N$^3$LL \cite{Catani:1989ne,Catani:2003zt,moch,magnea}. The impressive success of such a  program has brought up the idea that these processes can be used to monitor the parton luminosity at the LHC. This requires the theoretical uncertainty to meet the experimental one, which is thought to be at the percent level.

However, when the invariant mass $Q$ of the particles produced in the final state is
well below the centre-of-mass energy, the typical values of $x$ of the
colliding partons may be rather small: $x_1x_2 = Q^2/S\ll 1$.  Whenever $x$ is
small, logarithms of $x$ may spoil the perturbation series. Thus
accurate calculations require the computation of the coefficients
of these logarithms, and if they are large it may be necessary to resum them. This is particularly interesting for the LHCb experiment, which aims to measure the Drell-Yan production of a muon pair at invariant mass as low as $5$~GeV~\cite{LHCb}.

The resummation of small-$x$ logarithms in the perturbative evolution
of parton distribution functions at NLL has been performed by different groups
(see for example Ref.~\cite{heralhc} for a comparative review). Any reliable small-$x$ resummation should address the issue of the perturbative instability of the BFKL kernel; the resummed anomalous dimension should be computed at the NLL$x$ accuracy and must match standard DGLAP at moderate values of  $x$. In particular the resummation procedure proposed by Altarelli, Ball and Forte (ABF) provides a consistent treatment of the factorization schemes, which is an essential ingredient if one wants to match the resummation to fixed order calculations. For a description of the ABF resummation see  for instance~\cite{Altarelli:2008aj} and references therein.

The general procedure for resumming inclusive hard cross-sections at
the leading non-trivial order through $k_T$-factorization is known~\cite{CCH-photoprod,ch},
and its implementation when the coupling runs understood~\cite{ball}.
High-energy resummed coefficient functions are now known for an increasing number of processes.  Calculations have been performed for photoproduction processes
\cite{CCH-photoprod,ball}, deep inelastic processes \cite{Altarelli:2008aj,ch},
hadroproduction of heavy quarks \cite{CCH-photoprod,ellis-hq,camici-hq,ball},
and gluonic Higgs production both in the pointlike limit
\cite{Hautmann:2002tu}, and for finite top mass
\cite{Marzani:2008az,Marzani:2008ih}; the resummation for Drell-Yan and vector
boson production has been performed in~\cite{marzaniballDY} and, more recently, for direct photon production in~\cite{direct}.  Thanks to this effort is now possible to study the impact of small-$x$ resummation at the LHC.

\section{High-energy resummation}
\begin{figure}
\begin{center}
\includegraphics[width=0.4\textwidth]{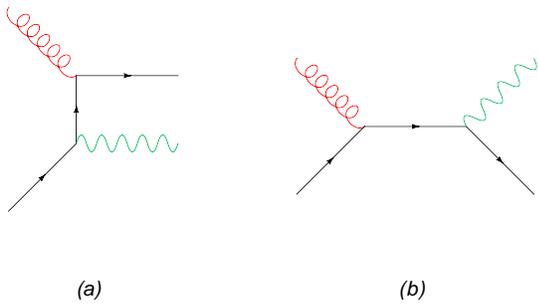}
\caption{Feynman diagrams for the process $ g^*(k) \;  q(p_1) \to  \gamma^* (q) \;q(p_2)$. The quarks are on-shell and massless, while the initial state gluon and the final state photon are off-shell.}\label{Fig:feyn}
\end{center}
\end{figure}

The leading non-trivial high energy singularities for inclusive cross-sections can be obtained from the $k_T$-factorization formalism \cite{CCH-photoprod},
and in particular its extension by Catani and Hautmann to deal with
situations in which the hard cross-section contains collinear
singularities which must be factorized consistently \cite{ch}. In \cite{marzaniballDY} we have extended such a procedure to deal with the rather complicated flavour structure which characterizes Drell-Yan processes. The main result of that computation is the resummed coefficient function for the quark-gluon sub-process. 
It can be computed by considering the process $$ g^*(k) \;  q(p_1) \to  \gamma^* (q) \;q(p_2) \,,$$ in the framework of  $k_T$-factorization. The relevant Feynman diagrams are shown in Fig.~\ref{Fig:feyn}, where $p_1^2=p_2^2=0$, but $k^2=-|\kt|^2$ and $q^2=Q^2$.

The analytic expression for the leading high energy behaviour of the coefficient function can be computed to any desired power of the strong coupling. Here we write it in Mellin space, where the limit $x \to 0$, corresponds to $N \to 0$; in $\msb$ we have:
\bea
 D_{qg}(N,\as)
&=& \frac{\as}{18 \pi} T_R \Big[1
+ \left(\frac{29}{6}
+{2 \pi^2}\right)\frac{C_A}{\pi}\frac{\as}{N}
\nonumber \\
 &+&\left(\frac{1069}{108}
+\frac{11}{3}\pi^2+{4}\zeta_3 \right)
 \left(\frac{C_A}{\pi}\frac{\as}{N}\right)^2\nonumber \\
 &+& \left(\frac{9031}{648}+\frac{85}{18}\pi^2+\frac{7}{20} \pi^4
+\frac{73}{3}\zeta_3 \right) \nonumber \\ &&
\left(\frac{C_A}{\pi}\frac{\as}{N}\right)^3 
 +\dots \Big]\,. 
\eea
The coefficients of $\ord \left(\as \right)$ and
$\ord \left(\as^2 \right)$ are in agreement with the
high energy limit of the fixed order NLO \cite{Altarelli:1978id}
and NNLO \cite{Hamberg:1990np,Blumlein:2005im} computations.
This is a very non-trivial check of the procedure. The
$\ord \left(\as^3 \right)$ and subsequent terms are all new results. The high energy singularities of the quark-quark coefficient function
are now easily deduced using the colour-charge relation:
\beq
 D_{qq}(N,\as)= \frac{C_F}{C_A}\left[D_{qg}(N,\as)-\frac{\as}{18 \pi} T_R\right]\,,
\eeq
which is valid in the high energy limit. It can be shown that all the other coefficient functions are sub-leading in the high energy limit~\cite{marzaniballDY}.
 \section{LHC phenomenology}
\begin{figure*}
\begin{center}
\includegraphics[width=0.3\textwidth, angle=-90]{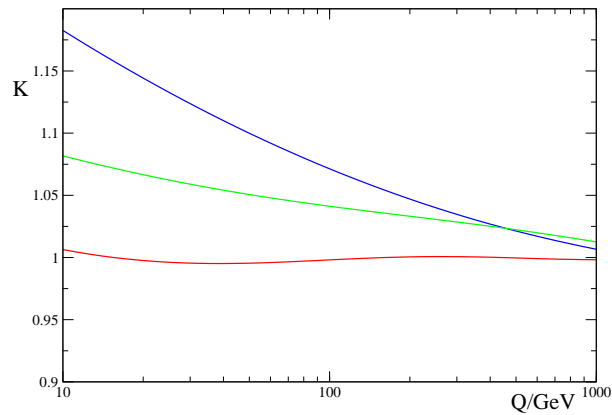}
\caption{In this plot from top to bottom we have: the NLO $K$-factor as defined in Eq.~(\ref{nlokfact}) and the two NNLO $K$-factors of Eq.~(\ref{nnlokfact}), $K^{NNLO}_1$ and $K^{NNLO}_2$ respectively. }\label{Fig:kfact}
\end{center}
\end{figure*}

We now to study the impact of high energy resummation on the inclusive Drell-Yan cross-section. We start by considering the resummed result matched to the fixed order calculation at $\ord(\as)$ using NLO parton distribution functions from the NNPDF collaboration~\cite{nnpdf}, with the evolution performed using the ABF resummed kernel. We define
\beq \label{nlokfact}
K^{NLO}= \frac{D_{ij}^{{\rm NLO }res} \otimes f_i^{{\rm NLO }res} \otimes f_j^{{\rm NLO }res} }{  D_{ij}^{{\rm NLO }} \otimes f_i^{{\rm NLO }} \otimes f_j^{{\rm NLO }}}\,,
\eeq
where $D^{\rm NLO}_{ij}$ are the coefficient functions for the different hard sub-processes and $f_i$'s are the parton distribution functions.
This $K$-factor is plotted in Fig.~\ref{Fig:kfact} as a function of Q (top line in blue) at the LHC centre of mass energy $\sqrt{s}=14$~TeV . At large $Q$ it approaches one, as it should;  at $Q=100$~GeV the resummation corrects the fixed order result by $7$~\% and the effect is as big as $15-20$~\% at $Q =10$~GeV. We then study the effect of the resummation on the NNLO calculation. Because the ABF resummation has not been implemented at this order, we consider two different ratios:
\bea \label{nnlokfact}
K^{NNLO}_1&=& \frac{D_{ij}^{{\rm NNLO }res} \otimes f_i^{{\rm NLO }res} \otimes f_j^{{\rm NLO }res} }{  D_{ij}^{{\rm NNLO }} \otimes f_i^{{\rm NLO }} \otimes f_j^{{\rm NLO }}}\, \nonumber \\
K^{NNLO}_2&=& \frac{D_{ij}^{{\rm NNLO }res} \otimes f_i^{{\rm NNLO }} \otimes f_j^{{\rm NNLO }} }{  D_{ij}^{{\rm NNLO }} \otimes f_i^{{\rm NNLO }} \otimes f_j^{{\rm NNLO }}}\,.
\eea
This definition of the $K$-factors is such that they again approach one at large $Q$; they are plotted in  Fig.~\ref{Fig:kfact},  $K^{NNLO}_1$  in green in the middle and $K^{NNLO}_2$ in red, bottom curve. 
The effect of the resummation is diminished. In particular $K^{NNLO}_2$ is always very close to one, suggesting the the NNLO coefficient functions capture most of the small-$x$ behaviour of the hard process.  In $K^{NNLO}_1$ instead the high energy contributions to the parton evolution is also taken into account. In this case the resummation corrects the fixed order results by $5-7$\% for $Q<100$~GeV. Thanks to these two results, we can estimate that the resummation corrects the NNLO by no more than a few percent for $Q<100$~GeV.

\begin{figure*}
\begin{center}
\vspace{0.5cm}
\includegraphics[width=0.3\textwidth, angle=-90]{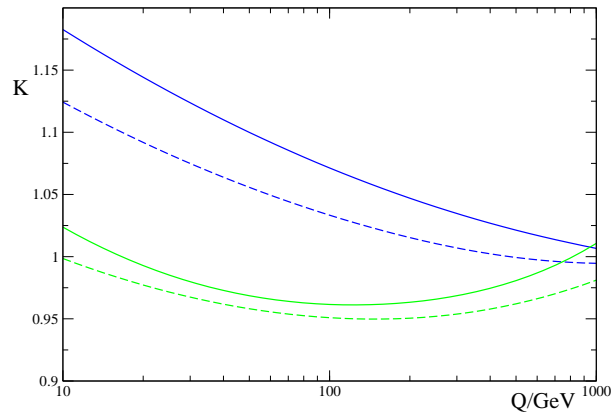}
\caption{In this plot we have the NLO  (top curves) and NNLO (bottom curves) resummed cross-section, normalised to NLO.}\label{Fig:kfactbis}
\end{center}
\end{figure*}

In Fig.~\ref{Fig:kfactbis} we compare the relative size of some of the various contributions beyond NLO. To this purpose we define four $K$-factors with the same denominator: the NLO cross-section computed with NLO partons. The two blue curves at the top are the NLO resummed cross-section with resummed partons (solid line, same as in Fig.~\ref{Fig:kfact}) and the NLO resummed cross-section with standard NLO partons (dashed line). The curves at the bottom are the NNLO resummed cross-section computed with NNLO parton distributions (solid green) and NLO ones (dashed green). 
The gluon-gluon subprocess makes a substantial contribution at NNLO. However this contribution has not been resummed yet: to do this would require the calculation of the off-shell process $g^* g^* \to \gamma^* q\bar{q}$.

\section{Conclusions}
We have presented analytic results for the Drell-Yan (and thus also vector boson production) 
coefficient functions to arbitrarily high orders in the strong coupling,  in the limit of high partonic centre-of-mass energy. Our results are given in $\msb$ scheme and they agree with the known results at NLO and 
NNLO, while providing new results at N$^3$LO and beyond.

We have evaluated the effect of high-energy resummation by computing K-factors between resummed and standard fixed order results. We have found that the resummation corrects the NLO result by $5$-$10$ \% for $Q< 100$~GeV. When NNLO corrections are included the effect is reduced to a few percent in the same kinematical region. 
We are currently working on the phenomenology for vector boson production.
 \vspace{0.3cm}
 
{\bf Acknowledgement}: we would like to thank Juan Rojo for providing us with suitable parametrizations of the NNPDF1.0 PDFs used in this study.

% ****************************************************************************
% BIBLIOGRAPHY AREA
% ****************************************************************************

\begin{footnotesize}
% IF YOU DO NOT USE BIBTEX, USE THE FOLLOWING SAMPLE SCHEME FOR THE REFERENCES
% ----------------------------------------------------------------------------

\end{footnotesize}

% ****************************************************************************
% END OF BIBLIOGRAPHY AREA
% ****************************************************************************

\end{document}